# Estimation of a diffusion model with trends taking into account the extremes. Application to temperature in France

G. Benmenzer[2], D. Dacunha-Castelle[1], T.T.Huong Hoang[1]

[1] Laboratoire de Mathématiques, Université Paris 11, Orsay, France
[2] Gaz de France Research and Development Division, Statistics and Artificial Intelligence Section

Correspondence to: T.T.Huong Hoang (huong.hoang@math.u-psud.fr)

**SUMMARY**. We built a model of the daily temperature based on a diffusion process and address to extreme values not taken in account into the literature on this kind of models. We first study, using non parametric tools, the trends on mean and variance. In a second step we estimate a stationary model first non parametrically and then using likelihood methods. Extreme values are taken into account in the estimation of model and to obtain a definitive estimation we use in a specific framework extreme theory for diffusions. A test of suitable model by simulation is done.

## INTRODUCTION

We would like to address the construction and estimation of diffusion models in a simulation perspective. Discrete models linked to diffusions probably remain until to day the best class of processes describing the daily temperature. They allow to compute by simulation the probability of some complicated events, often linked with the behavior of the diffusion process over high (low) thresholds. Recently, two fields have been more directly concerned with this problem: climate and finance, and often the two fields are concerned jointly with energy prices.

This paper is dedicated to a specific point: estimate and test a suitable model taking in account extreme values. Then apply the model to the daily temperature in France.

The starting point of the literature is a discretisation of an Ornstein-Uhlenbeck process using an Euler first order scheme, with a seasonal deterministic component, leading to the so-called Vacisek's model given by

$$dX_t = d\beta(t) + \alpha(\beta(t) - X_t)dt + \sigma(t)dW_t, \qquad (1)$$

where $\beta$ and $\sigma$ are deterministic seasonal components for the drift and the diffusion coefficient respectively, $\alpha$ allows the return to equilibrium for this stationary diffusion, W is a Brownian motion.

The discrete version is

$$X_n - \beta_n = (X_{n-1} - \beta_{n-1})(1-\alpha) + \sigma_n \varepsilon_n, \qquad (2)$$

where $\varepsilon_n$ is a standard Gaussian white noise.

Since these first works, (see Dornier et al. [15] for instance), some modifications or improvements have be done, (see Alaton et al. [2]). Brody et al. [7] argue for choosing a fractional Brownian motion in (1) instead of a Brownian motion, justified by long memory behavior but without statistical evidences. Benth et al. [3] take a Levy process (with a pure jump component) instead of the Brownian motion and work with hyperbolic distributions. Campbell et al. [8] propose important changes starting from the discrete version (2) where $\alpha X_{n-1}$ is replaced by $\sum_1^k \alpha_l X_{n-l}$ (the statistical procedure lead to *k = 3)*, so a longer memory and secondly $\varepsilon_n$ is no more a white noise but a GARCH process $\varepsilon_n^2 = \gamma \varepsilon_{n-1}^2 \eta_{n-1}^2 + \beta \eta_n^2$, where $\eta$ is a Gaussian white noise. We don't discuss here ARMA or more generally classical time series models, see Roustand [28], Moreno et al. [22] and Cao et al. [9] since they do not seem adapted to our problematic.

All these models take into account only what we call the "central part" component of the whole data when they test the quality of simulations, in fact they do not look to the extreme values, for instance they work only on quantile empirical intervals like $(q_\alpha, q_{1-\alpha})$, with $\alpha$ = *5/100*. Nothing is said about the quality of the model for tails and in our specific application for very high (or low) temperatures.

The previous models and more specifically the Vacisek's type one given by (1) has a good behavior in the central part when it is estimated using a learning period of observation of 50 years keeping one data per day to fix the ideas. But as we shall see in Part 1, it doesn't represent the extreme part, being to hot in winter and in summer. It is clear from a statistical study that there are various reasons to such an unsatisfactory behavior of the model (1).

The main reason lies in the choice of a diffusion coefficient which does not depend on the state $X_t$. As a consequence, they have let completely out the study of the extreme part values, for instance if the diffusion is stationary with an inaccessible boundaries $r_1, r_2$, they do not consider the question: are $r_1, r_2$ finite infinite; specific features on the estimated density of transition prove the inadequacy of models directly associated with an Ornstein-Uhlenbeck process.

Other basic problems are seasonality and non stationarity, and specifically trends produced in our example by an anthropic effect. It results from some works (see for instance Nogaj et al [24], Parey et al [26]) that the trends are not the same in extremes and in the central part, these trends are difficult to capture in the complete model. What is clear from the box plot data (see figure 6.1) is that the variability depends of course of time (seasonality) but also of state (temperature). For instance it increases strongly for low temperatures, physical evidence implies annual seasonality. So it is always easy to separate state and time effect for very low temperatures.

In order to split the technical problems, this first will address only problems which do not need a study of seasonality so we have divided our work into two papers. In the first one, centered on the general problematic, we will study data which is homogeneous with respect to the seasonality as for instance the data provided in winter. In a second paper centered on seasonality, we work on the whole data set which is more relevant for the applied results as it treats a much larger amount of data. Also some practical problems are induced in this first part for it remains some seasonality effects even in the chosen quite homogeneous seasons.

We start with a model

$$dX_t = b(X_t, t)dt + \sigma(X_t, t)dW_t \; , \qquad (2')$$

We choose a particular case non depending directly on t :

$$Z_{t+1} - Z_t = b(Z_t) + a(Z_t)\varepsilon_t \; . \qquad (2'')$$

where $Z_t$ is centered and normalized variable of $X_t$: $Z_t = \dfrac{X_n - m(t)}{s(t)}$ and *m(t)* and *s(t)* being respectively the mean and the variance trend. We discuss also the proper definition of trends.

We study the extremes of the process using the classical GEV theory. GEV theory gives useful arguments to choose the values of the inaccessible boundary ($r_1$, $r_2$) of the diffusion, $r_1$ can be taken as the lower bound of the distribution of $X_t$. Probability theory results (see Berman[5] and Davis[14]) allow to link GEV theory and the values of the drift and diffusion coefficients. It seems that the use of these tools is new in statistic and we have to make some weak hypothesis in order to built a bridge, convenient for a statistical use between extreme parameters estimation and diffusion coefficients ones. To estimate the model we follow these steps :

1-we estimate the trends m(t) and s(t)

2-Then we work with Z centered and normalized, $Z_t = \dfrac{X_n - m(t)}{s(t)}$

3-we estimate the values of the boundary of the diffusion and also the behavior of the drift *b* and the diffusion coefficient *a* near these boundary values

4-we use non parametric methods for a first estimate of *b* and *a*

5- we choose a two piecewise polynomial parametric model for *b* and *a* submitted to adaptative (estimated) constraints given by the first step: a has to be zero at the boundary and extreme theory fixes the slope of *a* at the same boundary. We also need to choose from data a specify value of first derivative of *a* at the break point. Under these constraints; the first polynomial of the is estimated using maximum likelihood, with a plug-in method, the other one is fixed by the constraints

An other way could be to use an approximated likelihood with respect to the size of the lag of discretization as introduced by one of the authors (Dacunha-Castelle et al [13]) and improved by Ait-Sahalia[1]. Computations are very heavy and so we do not use these ideas.

The series consist of 3-hours daily minimal temperature of about 50 years. We only work with the daily maximum in the hot season. So with respect to the diffusion, there is a method of measurement: maximization (minimization) and a discretisation. We don't discuss in this paper the effect of this two simultaneous operations (from theoretical point of view see Dacunha et al [13]) .We don't discuss the use of refined discretisation scheme, this point has to be treated in another work (see Talay [31]). Of course these problems are as important for extremes as for the central part for the lag of discretization plays a very important role (see Pitterbag[27]) for a discussion of this problem in the case of Gaussian processes).

The paper is organized as follows. In the second part we will estimate the mean trend and the scale trend of the series by the non parametric method. Part III is devoted to extreme theory. Practical procedures are proposes to include the results of extreme theory in the construction of the diffusion model. Part IV is devoted to the estimation of drift and diffusion coefficients. Non parametric and parametric methods are used and discussed. Part V and VI will be dedicated to simulation procedure. We test the validity and adequacy of the model using the simulation. To do that we will compare the estimates obtained from observation and from simulation for: densities (marginal and conditional), drift and diffusion coefficient, clusters length (defined as the number of consecutive days where the temperatures are larger (smaller) than a given threshold), GEV parameters. Finally in Part VII will be a discussion and will present some perspectives.

# 1. ESTIMATION OF MEAN FUNCTION AND SCALE FUNCTION

The mean *m* and the scale *s* are the function of *t* (the dates) ; they describe the behaviour of the temperature *X* for a large scale of time . The the drift and diffusion coefficients describe the variations of the temperature for a short scale of time. From a practical point of view and to check the trends effects, If we don't take account of the mean trend *m(t) and* scale trend *s(t),* that means we consider *m(t)* and *s(t)* as a constant, we have the formula:

$$X_{n+1} = X_n + b(X_n) + a(X_n)\varepsilon_n \qquad (3')$$

If only *m* is taken account, then we have:

$$Y_{n+1} = Y_n + b(Y_n) + a(Y_n)\varepsilon_n \qquad (3'')$$

$Y_n = X_n - m(n)$. If both of *m* and *s* are taken account: we come back to the formula (3).

In this section, we discuss the method to estimate *m* and *s*.

Moving average is often used to estimate the mean *m*. We have chosen as a non parametric method LOESS, or locally weighted scatterplot smoothing because of its intuitiveness and simplicity. This method has advantages to deal with the moving average. LOESS combines much of the simplicity of linear least squares regression with the flexibility of nonlinear regression. The fit at x is made using points in a neighbourhood of *x*, weighted by their distance to *x*. Throughout this paper, we use LOESS with a local polynomial of degree 1 (linear local smoothing).

The moving average gives a same weight to all points in the localized subsets, so part of the variation in the data is ignored. Let look at the figure 6.2 which describes the mean function estimator for the hot temperature. With a moving window of 1049 days, the estimator by LOESS has a similar behaviour than the estimator by moving average with a moving window of 501 days while the estimator of moving average with a window of 1049 days makes disappear some peak values.

Boundary effects are well-known in nonparametric methods; they affect the global performance and also the rate of convergence in the usual asymptotic analysis. The study of Ming-Yen Cheng and al [11] (1997) about the boundary effect in LOESS showed that a good solution to improve the estimation at the endpoints is just to apply the triangular weight kernel $K(u) = (1-|u|)I_{[0,1]}(u)$. In fact, for our data, even without this correction, the boundary effects of LOESS estimators don't create lot of troubles (see figure 6.2, 3).

The problem of robustness is antagonist to the use of information coming from extremes. Nevertheless, from a practical goal, we study the robustness for *m.* Cleveland [12] (1979) proposed a robust estimation procedure for LOESS by using Tukey's biweight. We can check the robustness of mean function *m* by this method. We remove respectively 1% to 5% of higher and lower values of the sample data and with each robust sample, we estimate *m* by LOESS and robust correction like above. We find that when we remove step by step extreme values, the mean behaviour doesn't change; they have the corresponding peak values at the same time; but there exists a constant distance between the estimators. So we use a very moderate 1% estimate.

. The smoothing parameter controls the smoothness and therefore has a crucial influence on the inference about *m*. There are several methodologies for automatic smoothing parameter as different cross-validation. But these criteria valid for i.i.d. data are too sensible to dependence to be use for our data without precaution. So we have to use often selection like rules of thumb or the minimization of the global for prediction to select a suitable parameter which allows to obtain a bias and variance trade-off.

Variance function estimation in regression is more typically based on the squared residuals from a preliminary estimator of the mean function. After estimating mean trend *m*, we can estimate the scale function when we consider *s* like the mean function of square residuals regression. So the estimator of *s* has a form: $s^2(x) = \sum_i w_i(x)(X_i - \hat{m}_i(x))^2$ where w is the chosen weight function. Recently Ruppert [29] used the linear local regression for variance estimation; the influence of estimator of the mean function and the selection of bandwidth is also discussed. Ruppert has shown that the leading bias and variance term for our local polynomial variance estimation are analogous to those for local polynomial estimator of the mean function; so that there is no loss in asymptotic efficiency due to estimating *m*. Ruppert proposed the same bandwidth selection for both of estimations of mean function and scale function.

## 2. EXTREME THEORY IN DIFFUSION PROCESS

In order to apply GEV theory to diffusions, let us state some results from probability theory. The first one concerns the maximum on an interval for a stationary diffusion when the length of the interval increases. The basic results are due to Berman [5] but the version of Davis [14] fits a statistical framework better. First, we are concerned by a stationary diffusion with invariant density $v$ and inaccessible boundaries $r_i$ and $r_s$ which are finite or infinite but unaccessible ( Breiman [6]).

Let a diffusion *X* with liptchitzian coefficients a and *b* defined by :

$$dX_t = b(X_t)dt + a(X_t)dW_t$$

the scale function *s* of the diffusion ( ) is given by $s = \frac{1}{2}\int^x e^{-\int^u 2\frac{b(v)}{a(v)}dv} du$ (we keep the traditional notation s for there is no possible confusion in the paper with the trend s of the variance)

**Theorem** *If $M_T$ = max($X_t$), 0 < t < T, if it exists functions A and B such that $\frac{M_T - B_T}{A_T}$ converges in distribution to G, then G is a GEV distribution define in the following manner : if F is the distribution defined by then F is in the (extreme) domain of attraction of G.*

**Remark** : This means that the GEV distribution G associated to the max value in a large block of consecutive observations at discrete times is not in general the same GEV distribution H associated to a sample of the marginal density $v$ . It can success that the domain of attraction of H is different of G even for instance with different shape parameters ( see Davis (9) for an example ).

**Application** : we can see from the previous theorem that F has particular properties : as a and *b* are continuous on the interval ($r_I$, $r_S$) (and even in general supposed liptchizian) functions with a>0 we see that F is twice continuously differentiable, so the formula giving F shows that

$$\frac{s''}{s'} = \frac{-2b}{a} = -\frac{F'}{F}\left(1 + \frac{2}{\log(F)}\right) + \frac{F''}{F'} \quad (**)$$

from *s* tends to infinity as *X* tends to $r_l$ or $r_S$, and in this case a tends to 0 . Suppose that F is in the domain of max-attraction for some GEV distribution G with shape parameter ξ<0. We now from the general theory of extremes (Embrechts et al[17]) that this means that as *x* tends to $r_S$, we have 1- F(x) equivalent to $L(x)x^{-1/\xi}$ so this means first that the upper bound of g is $r_S$ and that 1/*s* is equivalent to $L(x)x^{-1/\xi}$ as *x* tends to $r_S$.

**Definition** *We say that a $C^2$-function f with regular variation is of $C^2$-regular variation as x tends to $r_S$ if for some ξ, we have f''(x) equivalent to $x^\xi L(x)$ with L of slow variation as x tends to $r_S$ . For instance, if 1-F(x) is equivalent to to $x^\xi L(x)$ with L is a function product of different powers of iterated from different orders logarithms then 1-F(x) is of $C^2$-regular variation.*

From a statistical point of view, it is well known that it is very difficult (quite impossible!) to estimate L when ξ is unknown and even if it is known ( as it is very difficult to know if F is in a max domain of attraction). So to suppose for a twice differentiable function that it has the $C^2$-property of regular variation it is not in practice restrictive with respect to the hypothesis"it belongs to a max-domain of attraction".

So we can obtain the following lemma:

**Lemma**: *Suppose F of $C^2$-regular variation and in the domain of attraction (for maximum) of $G(\mu, \sigma, \xi), \xi < 0$, let $r_S$ the common upper bound of F and G. Then we have the previous behaviour (**) for F as X tends to $r_S$, this result is independent of the slowly varying function L (which can be of course constant)*

*We have the following behavior of a near the upper bound $r_s$, $a(x) \approx -2b(x)(r_s - x)(-\frac{1}{\xi} - 1)$*

This lemma thus proves that if two diffusions have the same drift and the same $\xi < 0$ and the same upper bound then necessarily near this bound the diffusion coefficient are equivalent to the same linear form, in fact they are linear up to the first order and are also the same for every $(\mu, \sigma)$ such that $r_S = \mu - \frac{\sigma}{\xi}$ as $\xi$ are fixed.

**Proof** : from $F''(x) \propto L(x)x^{(-\frac{1}{\xi} - 2)}$ we deduce by integration the equivalences for F' and log F and apply the formula (1).

So we now have an estimate $\hat{a}$ by the plug in method of a near the bounds if, $\xi < 0$ for the two bounds, $\hat{a} = -2\hat{b}(x)(\hat{r}_s - x)(-\frac{1}{\xi})$

We only consider this case $\xi < 0$ here, for temperatures in France have negative $\xi$ .

## 3. ESTIMATION OF THE DRIFT AND THE DIFFUSION COEFFICIENT

Estimation of the diffusion process or stochastic differential equation has been considered in the literature for many years, with some the papers being concerned with estimating the drift and diffusion functions from continuously sampled data. However with discretely sampled observations from the continuous sampling path, estimation of the continuous-time diffusion process is more complicated and more if one has to discuss the role of the lag for discretization. We work with nonparametric estimates first and then with parametric estimates for the drift and diffusion coefficient.

The first parametric estimator of the coefficients of a stationary diffusion process from discretely sampled observations taking in account the lag is the approximate maximum likelihood estimator proposed by Dacunha-Castelle and al [13] (1986). This result was improved by Ait- Sahalia[1]. Other parametric estimators include the maximum likelihood estimators derived by Lo [20](1988) for more general jump-diffusion processes, the method of moments based on simulated sampling paths from given parameter values proposed by Duffie and Singleton [16](1993). Many authors [30](1997) used kernel regression or wavelets for instance to construct nonparametric estimates for the diffusion coefficients .

We first estimate *b* in the supplied model by local smoothing LOESS where *b* is considered as the conditional expectation $E(Z_n-Z_{n-1}|Z_{n-1})$. The behavior of $\hat{b}$ in the central part from 1%-quantile to 99%-quantile is rather linear. Local estimation in 1% higher and lower value, because of lack of observations, seems to have no physical and probably statistical meaning using this method. Besides, when considering *b* like a polynomial and using likelihood tests, we can find out a polynomial of degree 1 is the optimal model for *b*. Comparing local estimator $\hat{b}$ with linear least squares estimator of *b*, we find that they merge in the central part (see figure 6.3). In summary, we can consider *b* as linear.

A study on the robustness of *b* is necessary. It comes back to the same idea in the previous section: that is to remove respectively 1%, 2%, …,5% higher and lower of the sample and observe how the drift estimator changes in the central part. The linearity of *b* follows acceptable by likelihood tests. We obtain always a negative linear drift but more the values in extreme parts are removed, steeper is the slope of the drift. For example, with a stationary diffusion model for the cold season, from a slope of drift β= -0.22 with nothing removed, we have a slope β=-0.29 with 5% higher and lower values removed. This gap is rather large. We know that the estimators by least squares are not robust. Indeed this change is more clearly seen for cold temperature. The cold temperature is less homogenous than the hot temperature (see figure 6.1). For this reason, if we use a robust *b*, it improves the results in the central part but gives a bad estimation for high quantiles  For we are interested in extreme values, we don't use robust estimator; we keep all the observations sequence for estimating *b*.

To estimate *a*, we can consider it as the square root of the conditional variance function of $Z_n$ given $Z_{n-1}$. We estimate it using a kernel method. If *K* is a kernel (Epanichkov or Gaussian are used) and *h* is the corresponding bandwidth then the conditional density is estimate as

$$\frac{\sum_{n=1}^{N} \frac{1}{h_N^2} K\left(\frac{Z_n - x}{h_n}\right) K\left(\frac{Z_{n-1} - y}{h_n}\right)}{\sum_{n=1}^{N} \frac{1}{h_N} K\left(\frac{Z_{n-1} - y}{h_n}\right)}$$

and its mean and variance as usually directly, the summand in the numerator being multiplied by an adequate term.

Then, we compare the results with the estimate obtained from a penalized likelihood method . We start from the density

$$\frac{1}{a(Z_{n-1})} \exp\left\{-\frac{1}{2}\left(\frac{Z_n - Z_{n-1} - b(Z_{n-1})}{a(Z_{n-1})}\right)^2\right\}$$

So the penalised likelihood is given

$$L(n,a) = \sum_{i=2}^{n}\left[-\frac{1}{2}\left(\frac{Z_i - Z_{i-1} - \hat{b}(Z_{i-1})}{a(Z_{i-1})}\right)^2 - \log a(Z_{i-1})\right] - \frac{1}{2}\lambda \int [a''(x)]^2 dx \qquad (4)$$

with a convenient regularization parameter $\lambda$, where $\hat{b}$ is estimated by linear least-squares: $\hat{b}(Z_{n-1}) = \alpha + \beta Z_{n-1}$ as previously. The maximization of l gives $a$ as a cubic spline function. This is the first step. A recursive procedure checks the stability of estimation of $a$ and $b$. That means with $\hat{a}$ estimated from (4) (called original $\hat{a}$), we estimate a new linear $b$ by maximizing:

$$\sum_{i=2}^{n}\left[-\frac{1}{2}\left(\frac{Z_i - Z_{i-1} - b(Z_{i-1})}{\hat{a}(Z_{i-1})}\right)^2 - \log \hat{a}(Z_{i-1})\right]$$

and we resume. The result of these work iterations give us the same estimators of $a$ and $b$. In deed, we can also estimate $a$ and $b$ at the same time by (4), this is theoretically possible but the numeric implementation takes too much of time.

We pass now to a parametric estimation, the parametric model is suggested by all the previous results, the one on extremes and the non parametric estimates
We estimate $a$ and $b$ at the same time first by considering $a$ as a polynomial p (of degree 4 in our case):

$$\sum_{i=2}^{n}\left[-\frac{1}{2}\left\{\frac{Z_i - Z_{i-1} - b(Z_{n-1})}{p(Z_{i-1})}\right\}^2 - \log p(Z_{n-1})\right] \qquad (4')$$

So after maximizing (4'), we obtain estimators of parameters: 2 for $b$ and 5 for $a$.
We can compare the behavior of $\hat{a}$ obtained by parametric and non parametric methods. Near the boundary, say 1% higher and lower values of $X$, estimators of a and b have not sense for there are not enough date to perform directly any estimate and so we have to use extreme theory to make a kind of extrapolation. So the comparison excludes this neighborhood of the boundary. The parametric estimator of $a$ in (4') is retained for implementation. With these parametric estimators of $a$ and $b$, we obtain the residuals which can be tested. They are accepted as a white noise after considering their first and second moments, their autocorrelation plot and qq-norm plot.
Next a same work to $b$ is realized for checking the robustness of $a$. But in this case, we focus more on the extreme part. We remove just respectively 1% to 3% of lower values.

From the three robust samples, we consider the difference of the estimated $a$ from robust samples with one of original sequence: $\Delta(X,Y) = \sum |\theta_{X,i} - \theta_{Y,i}|$ with $\theta$ the coefficients of polynomial of 4 degree, estimator of a. We obtain the values of $\Delta$ of 3 robust samples with the whole series respectively: 0.021, 0.027 and 0.05. The third one(-3%) has different behavior of $\hat{a}$ in some parts. In the first one (-1%) and the second one (-2%), the difference can be acceptable.

We propose now to separate the estimation of a and b in the central part (from 1%-quantile to 99%-quantile) and extreme part (from 1% to the boundary): $a_{cent}$ and $b$ are estimated by all of the observations. Then $a$ in the extreme part (>99% and <1%) is estimated differently.

The drift function has a quite clear physical meaning as the elastic part of the basic oscillator which the deterministic justification of the Ornstein –Uhlenbeck first model. So we suppose that this drift is valid until the boundary and we discuss later this hypothesis.

When applying GEV distribution for the maxima (respectively minima) of our series of temperature in France, usually we obtain $\xi<0$, which means the temperature is bounded. And their upper bound (respectively lower bound) can be estimated from a function of the estimators of extreme parameters of GEV distribution: $r_S = \mu - \dfrac{\sigma}{\xi}$

Testing the quality of the estimators of *a* and *b* in formula (4') by simulation, we can see that the simulation-based empirical quantiles (from 1000 samples) in the central part (from 5% to 95% quantile) are really close to the true quantiles : a maximal difference of 0.5°C is found. However in the tail, (from 1% to 3%)we find out a difference of about 1.5°C. This bias in climatology is rather large even if we take into account that the record observed on 1000 samples is much larger than the observed one and so this has an influence on the 1% quantile. , In summary, simulated tails are heavy than the observed quantile as a simple consequence of the law of large numbers. It remains to give a more precise definition of "where the tail begins" in the whole distribution.

We need a correction for the diffusion coefficient *a* in the tail. Our idea is based on a theoretical result, the behavior of *a* at the boundary  In the previous section, we have applied GEV distribution for the extreme values of diffusion process and we obtained a parabolic form in the neighborhood of the boundary (see lemma, part 3). There is not any assurance that this is well modeled for all "extreme" part. So for another part we look for a suitable experimental estimator of *a* for 1% higher and lower values.

Thus the lack of theoretical results obliges to use an interpolation in order to complete the function $\hat{a}$. We look for here a suitable polynomial of degree 2 (to follow to the extremes theory) which satisfies the condition at the boundary *(a(r$_S$) =0)* and two another constraints are added that $a_{ext}(q_{0.01})=a_{cent}(q_{0.01})$ and $a'_{ext}(q_{0.01})$ given by the lemma. In another tail, higher than 99%-quantile for the cold temperature, do we need a correction of *a?* The high values in daily minimum temperature in the winter are not really homogenous and in fact their role is not really important, we can use $a_{cent}$ for this part.

*Remark:* These estimators are then built by a plug-in method from the estimators given by GEV estimates and by an empirical quantile, the remaining is provided by the likelihood method. So it can be show that the estimators of *a* and *b* are consistent, with a speed of convergence $1/\sqrt{n}$, the computation of the matrix of information is very complicated and details will be given elsewhere.

## 4.   SIMULATION AND RESULTS

We simulate 3 models:

1/  *m* and *s* are considered as constant:

2/  *m* is a function of time and *s* is constant

3/ both of *m* and *s* are functions of time.

In order to validate a suitable model for our data, we make 1000 simulations basing on the models above for the low temperature.

To test the quality of the model, we have to determine whether the simulation model is an accurate representation of the observations. The comparison is based on the following items:

- ❏ Quantiles

- Marginal density, mode, mean, variance
- GEV estimated distribution and its parameters
- Clusters distribution: number of clusters and length for observation less than the threshold given by quantiles 1% and 2% and larger than threshold given by quantiles 98% and 99%

**The simulation procedure:**

MODEL 1:

- Estimate *a* and *b* from the initial sequence *X* with the method described in part 4
- With these estimated *a* and *b*, simulate 1000 samples with a recurrence procedure by using $X_{n+1} = X_n + b(X_n) + a(X_n)\varepsilon_n$

MODEL 2:

- Estimate the mean function *m* of *X* and calculate Y=X-m
- Then use the same procedure as for model 1 with Y instead of X
- Deduce 1000 samples of *X* by adding *m*: X=Y+m

MODEL 3:

- Estimate the mean function *m* and scale function *s* of X, calculate Z=(X-m)/s
- Then use the same procedure as for model 1 with Z instead of X
- Deduce 1000 samples of *X* by adding *m* and multiplying *s*: X=sZ+m

We obtain the results for cold temperature:

❖ *Quantiles:*

Below we construct a table with the empirical quantiles corresponding to three kinds of models for the cold temperature. We consider from 1st percentile to 99th percentile, smaller or higher percentiles have not a significant meaning. Let O for observation and S for simulation

| Percentiles | O | S without *m* and *s* | S with *m* | S with *m* and *s* |
|---|---|---|---|---|
| 1% | -14.40 | -14.07 | -14.96 | -16.21 |
| 2% | -12.44 | -11.78 | -12.27 | -12.74 |
| 3% | -11.40 | -10.43 | -10.79 | -11.08 |
| 5% | -9.40 | -8.73 | -8.98 | -9.13 |
| 10% | -6.60 | -6.42 | -6.58 | -6.64 |
| 30% | -2.20 | -2.49 | -2.53 | -2.55 |
| 50% | -0.10 | -0.18 | -0.19 | -0.20 |
| 70% | 1.90 | 1.98 | 1.96 | 1.95 |
| 80% | 3.20 | 3.26 | 3.23 | 3.21 |
| 90% | 5.30 | 5.08 | 5.00 | 4.97 |
| 95% | 6.80 | 6.66 | 6.54 | 6.50 |
| 97% | 7.51 | 7.77 | 7.61 | 7.57 |
| 98% | 8.14 | 8.63 | 8.46 | 8.42 |
| 99% | 9.30 | 10.11 | 9.94 | 9.92 |

Three cases give us the different results of empirical quantiles, especially in the cold extremes. In the central part from 1st percentile to 95th percentile, we don't find an important difference between the three models. A difference about 0.1°C is found for the median, but this difference is smaller for the percentile s close to the median. This is not found for the hot temperature but for cold temperature it is implied by a remaining of seasonality which as we will show in the second paper disappears when the whole data is taken in account.. However 0.1 of difference is not really significant. The empirical quantiles for three models can be considered rather good versus the true quantiles.

In the extreme part, from 95th to 99th percentiles, the third model with *m* and *s* seems a little better than other ones. For 1% and 2% quantiles, the third model seems better with the difference larger for the smaller percentiles, while which is inverse for two other ones.

In the case of hot temperature, it is surprising that the results in 1st model and 2nd model with mean trend are mostly analogous. *m* in summer has a total variability of 3°C, that means that *m* cannot be considered as constant. The diffusion cannot be considered as stationary. Indeed, observing two models: stationary and stationary with a mean trend, we can see that when *m* is difference-stationary ($\Delta m$ is stationary) with zero mean, *b'(X-m)* plays the role of mean function of $\Delta X$, so it is identical with *b(X)*. From the same argument, we have *a(X)=a'(X-m)* , where *b, b'* and *a, a'* are respectively drift functions and diffusion functions of stationary and non-stationary diffusion process.

Hot temperature has a linear trend for the mean. With a ADF test, we can check it. In the figure 6.5, we can find out the similarity of *a* and *b* in two models. No difference can be seen between the empirical quantiles because the quantiles don't pay attention to the order of values.

*Marginal density:*

The simulations of all three models have a median and mean near by the true median and true mean. However their variance is smaller; their values are more concentrated near the mode which is also the median. We already explain this defect by the remaining seasonality

❖ *Estimators of GEV parameters*

With our modelling in the tail of *a*, the empirical estimators of GEV parameters are rather good (see figure 6.5). The mean of the simulated location parameter $\mu$ approaches the true $\mu$. The scale parameter $\sigma$ is not so well estimated for the hot temperature because the simulations can catch of the strong variability of the extreme hot temperature in the last years. The shape parameter $\xi$ is difficult to estimate, but in our case, the simulated empirical estimation of $\xi$ is very good.

❖ *Clusters*

We concentrate on the rate of declusterization (which is can be considered as the inversion of average length of the clusters) and the distribution of clusters.

For hot temperature, the average length of clusters of simulations is smaller than the observations one. The distribution in the 2nd simulated model (with *m*) seems to be nearest to that of observations.

For cold temperature, the results of the 3 models are rather similar.

# 5. GRAPHICS

## 5.1 Data estimations

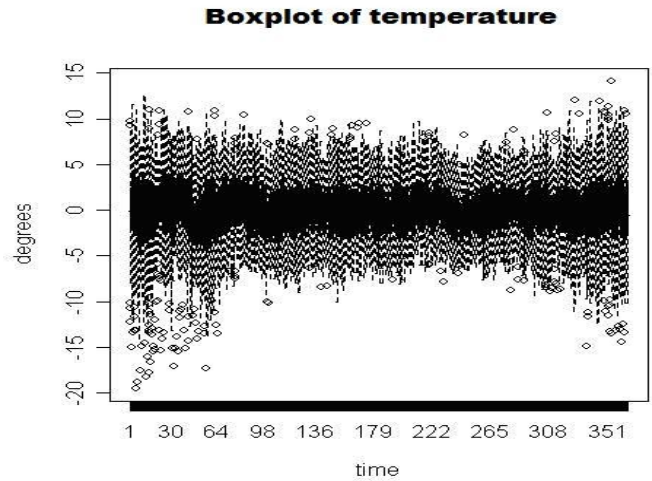

Box plot of data without seasonal component for the mean shows the seasonality of the mean and of the variance, this one is very large in Winter and the marginal densities are asymmetric

## 5.2 Advantage of LOESS method versus moving average:

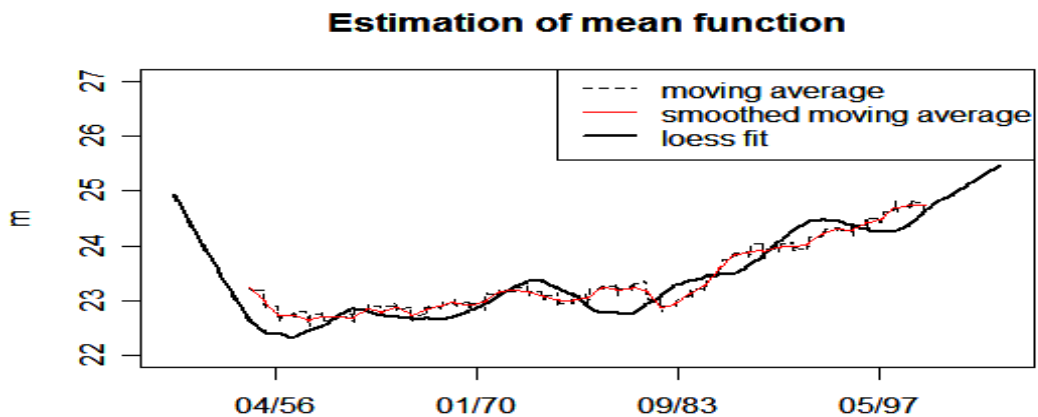

The mean trend estimators by moving average and LOESS with a window of 1049 days (correspond to 11 years)

## 5.3 Estimation of drift function

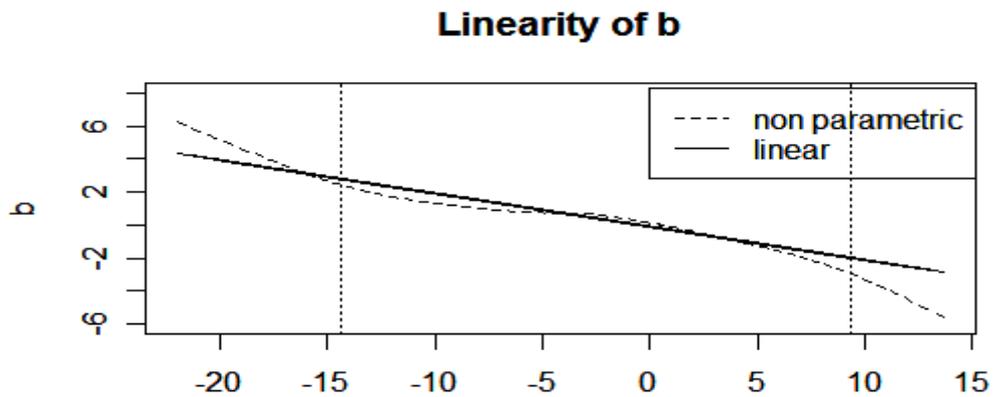

*b* estimated for the sequence of temperature in cold season in Strasbourg

*b* estimated by LOESS approach b estimated by linear least square in the central part
(which is limited by the vertical lines)

## 5.4 Estimation of diffusion coefficient a

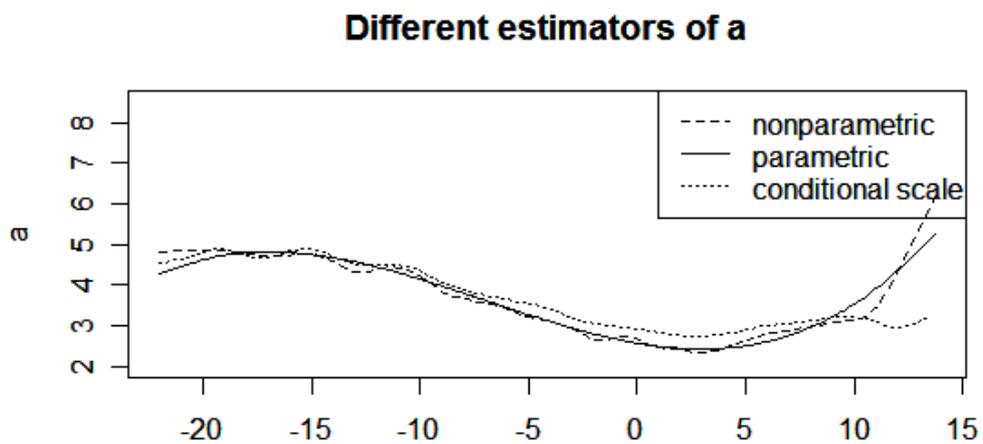

Different estimations of *a* in the central part : non parametric and parametric
The oscillations are not significative (lack of data) for very low temperatures

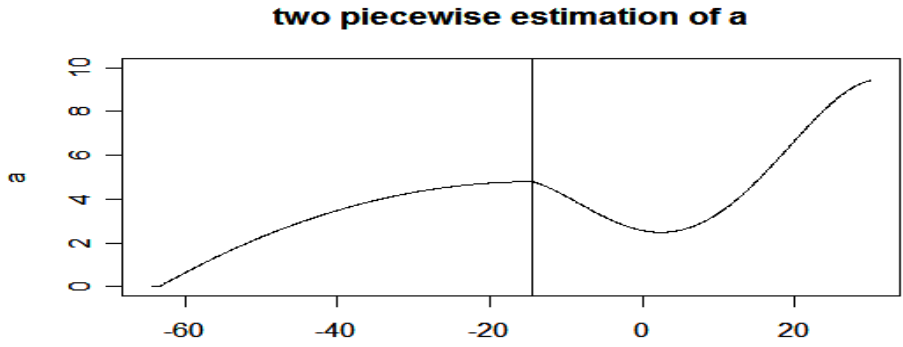

Estimation of a in the central part and extreme part which is separated by the dash-line of the 1st percentile

## 5.5 Simulation

1000 samples of simulation for the diffusion equation.

*Diffusion coefficients*

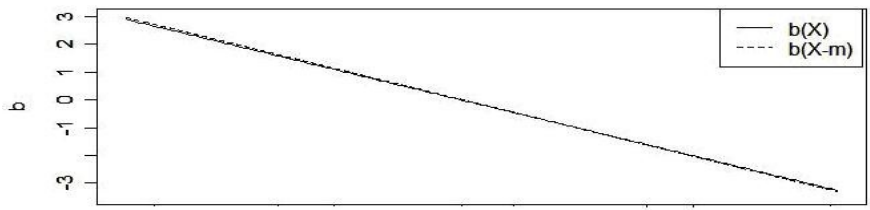

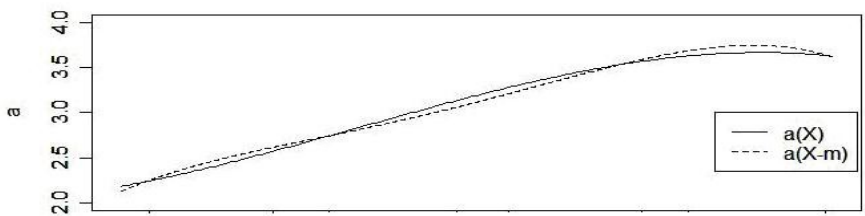

Comparison between a(X), b(X) in 1st model and a'(X-m), b'(X-m) in 2nd model with mean trend: they are mostly coinciding

*Marginal density*

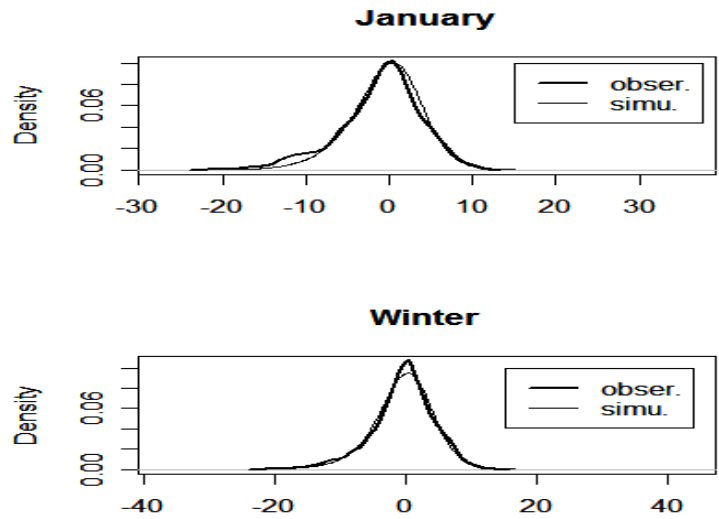

Marginal density of the observations and the simulations

The effect of seasonality remains in the simulated model of the winter (3 months). It implies that the variance of the simulations is smaller than that of observations. This fact is not found in the simulated model for just one month (January) but a clear difference is found in the cold extreme part because of the lack of values.

*GEV parameters*

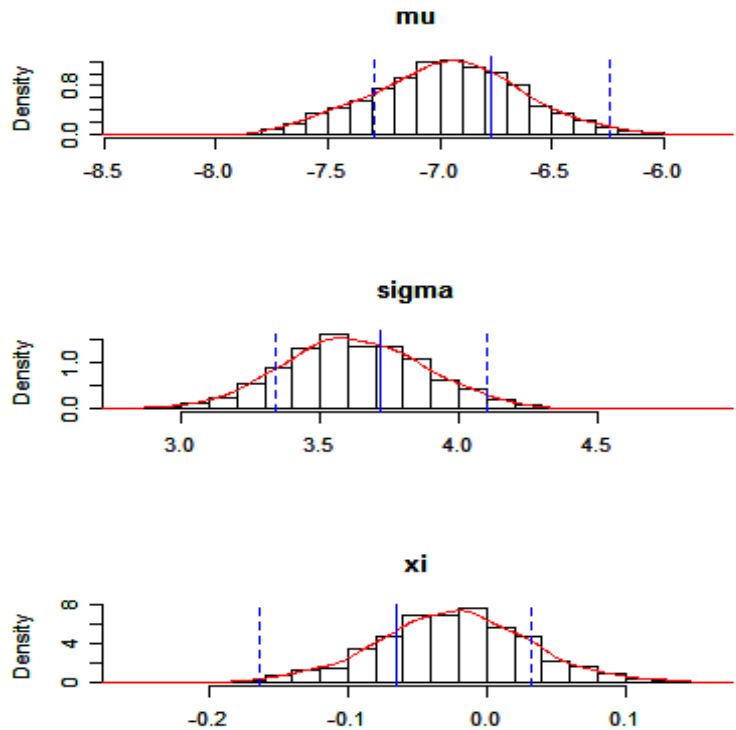

Estimation of GEV parameters in the simulations

The vertical lines are the true parameters with their confidence intervals of 90%

*Clusters*

❖ Hot temperature

|  | *O* | *S 1$^{st}$ model* | *S 2$^{nd}$ model* | *S 3$^{th}$ model* |
|---|---|---|---|---|
|  |  | **Quantile 98%** |  |  |
| **Threshold** | 32.80 | 33.15 | 33.11 | 33.19 |
| Length | Distribution | Distribution | Distribution | Distribution |
| 1 | 0.700 | 0.695 | 0.715 | 0.703 |
| 2 | 0.150 | 0.202 | 0.193 | 0.195 |
| 3 | 0.050 | 0.070 | 0.061 | 0.065 |
| 4 | 0.033 | 0.022 | 0.020 | 0.024 |
| 5 |  | 0.007 | 0.007 | 0.008 |
| 6 |  | 0.003 | 0.003 | 0.003 |
| 7 |  | 0.001 | 0.001 | 0.001 |
| **Rate of declusterization** | 0.588 | 0.686 | 0.703 | 0.686 |
|  |  | **Quantile 99%** |  |  |
| **Threshold** | 34.00 | 34.47 | 34.43 | 34.60 |
| **Clusters** Length | Distribution | Distribution | Distribution | Distribution |
| 1 | 0.824 | 0.776 | 0.807 | 0.778 |
| 2 | 0.059 | 0.174 | 0.148 | 0.161 |
| 3 | 0.088 | 0.039 | 0.033 | 0.043 |
| 4 | 0 | 0.009 | 0.009 | 0.013 |
| 5 |  | 0.001 | 0.003 | 0.004 |
| 6 |  |  | 0.001 | 0.001 |
| **Rate of declusterizatio** | 0.654 | 0.777 | 0.796 | 0.764 |

❖ Cold temperature

|  | *O* | *S 1$^{st}$ model* | *S 2$^{nd}$ model* | *S 3$^{th}$ model* |
|---|---|---|---|---|
|  |  | **Quantile 2%** |  |  |
| **Threshold** | -12.44 | -11.77 | -11.77 | -11.76 |
| **Clusters** Length | Distribution | Distribution | Distribution | Distribution |
| 1 | 0.478 | 0.564 | 0.568 | 0.571 |
| 2 | 0.283 | 0.203 | 0.205 | 0.201 |
| 3 | 0.065 | 0.099 | 0.097 | 0.098 |
| 4 | 0.022 | 0.053 | 0.054 | 0.053 |
| 5 | 0.043 | 0.032 | 0.030 | 0.032 |
| 6 | 0.022 | 0.018 | 0.018 | 0.019 |
| 7 | 0 | 0.011 | 0.011 | 0.011 |
| 8 | 0 | 0.007 | 0.007 | 0.007 |
| 9 | 0.022 | 0.005 | 0.004 | 0.004 |
| 10 | | 0.003 | 0.003 | 0.002 |
| 11 | | 0.002 | 0.002 | 0.001 |
| 12 | | 0.001 | 0.001 | 0.001 |
| 13 | | 0.001 | 0.001 | |
| **Rate of delusterization** | 0.447 | 0.497 | 0.505 | 0.505 |
| | **Quantile 1%** | | | |
| **Threshold** | -14.4 | -14.07 | -13.98 | -13.97 |
| **Clusters Length** | Distribution | Distribution | Distribution | Distribution |
| 1 | 0.600 | 0.588 | 0.597 | 0.599 |
| 2 | 0.167 | 0.204 | 0.206 | 0.201 |
| 3 | 0.200 | 0.098 | 0.093 | 0.095 |
| 4 | 0 | 0.049 | 0.047 | 0.049 |
| 5 | 0.033 | 0.027 | 0.026 | 0.026 |
| 6 | | 0.015 | 0.013 | 0.013 |
| 7 | | 0.008 | 0.008 | 0.008 |
| 8 | | 0.005 | 0.004 | 0.004 |
| 9 | | 0.003 | 0.003 | 0.002 |
| 10 | | 0.001 | 0.001 | 0.001 |
| 11 | | 0.001 | 0.001 | |
| 12 | | 0.001 | | |
| **Rate of declusteriztio** | 0.588 | 0.535 | 0.547 | 0.546 |

## CONCLUSIONS AND PERSPECTIVES

Our purpose is to find out a parametric diffusion. We choose the model which takes in account the mean trend and scale trend. The most interesting result is that the modification of the diffusion coefficient now highly non linear as a function of the temperature gives a model of simulation with the following properties:

1- It keeps the qualities of the classical seasonal model based on Ornstein-Uhlenbeck model for data in the central part

2- At the opposite of the classical model, it gives a good fit to the data (observed quantile q) for 2%<q<5% and the same for large values, 2% at least.

3- The 1% quantile estimated by simulation seems physically acceptable, we have no manner to estimate it from data, we use extreme theory and the results agree. Extreme distribution parameters are rightly simulated.

Of course nothing can be test for quantiles less than 1%, for instance the bimodality phenomena of conditional densities for high absolute values which seems possible is recovered on simulations but not clear for observations for we have no sufficiently data

**Perspectives**

1- As we say previously, seasonality will be taken in account in a former paper; it shall give better results for the data are much more important despite the necessity to introduce new parameters

We try some ways of improvement of the methodology:

We can change the discretisation scheme, use a refined form as Milshtein one [21]. And add a short memory to the model, increasing the degree of the autoregressive part

2-We observe as we say bimodality of the transition density for very high or very low values. This is and interesting point from climate point of view, for these values it seems that often the return to equilibrium is balanced by a persistence phenomena

To take into account the more accurate statistical analysis and specifically the bimodality of the transition density, it is perhaps necessary to change the discrete scheme. Many ways can be explored in this direction.

To think the problem in a simplest way with a physical interpretation (in term of weather types in the application) and take for $\varepsilon_n$ a white noise with a conditional distribution given by a mixture of Gaussian noises.

$$p(X_{t-1})\exp\left\{-\frac{1}{2}\left(\frac{x - m_1(X_{t-1})}{\sigma_1(X_{t-1})}\right)^2\right\} + (1 - p(X_{t-1}))\exp\left\{-\frac{1}{2}\left(\frac{x - m_2(X_{t-1})}{\sigma_2(X_{t-1})}\right)^2\right\},$$

with $p(X_{t-1}) = 1$ for $X_{t-1} \leq S_1$ or $X_{t-1} \geq S_2$, $S_1 < S_2$ convenient thresholds.